\documentclass[a4paper]{PoS}
\pdfoutput=1
\usepackage[T1]{fontenc}
\usepackage[utf8]{inputenc}
\usepackage{slashed}
\usepackage{amsmath}
\usepackage{textcomp}

\title{Numerical studies of Minimally Doubled Fermions
}

\ShortTitle{Numerical studies of Minimally Doubled Fermions}

\author{\speaker{Johannes Heinrich Weber}\\
        Johannes Gutenberg-Universit\"at Mainz,
        Institute for Nuclear Physics
        \\
        University of Tsukuba,
        Graduate School of Pure and Applied Sciences
        \\
        E-mail: \email{weberj@kph.uni-mainz.de}}

\author{Stefano Capitani\\
        Johannes Gutenberg-Universit\"at Mainz,
        Institute for Nuclear Physics
        \\
        E-mail: \email{capitan@kph.uni-mainz.de}}

\author{Hartmut Wittig\\
        Johannes Gutenberg-Universit\"at Mainz,
        Institute for Nuclear Physics
        \\
        E-mail: \email{wittig@kph.uni-mainz.de}}

\abstract{We have performed the first numerical study of minimally doubled fermions of the Karsten-Wilczek class in the quenched approximation. This requires fixing the counterterms, which arise due to hypercubic symmetry breaking induced by the Karsten-Wilczek term.
Non-perturbative renormalisation criteria are formulated after a detailed study of the parameter dependence of mesonic observables. Minimisation of the mass anisotropy of the pseudoscalar ground state fixes non-perturbative renormalisation conditions for the counterterm coefficients. These anisotropies are mapped out by probing different euclidean components of the transfer matrix through calculations of the pseudoscalar ground state mass in different directions.
The chiral behaviour of the pseudoscalar ground state is studied with the tuned Karsten-Wilczek action for multiple lattice spacings. Light pseudoscalar masses ($ M_{PS} \lesssim 250\,MeV $) were achieved in the quenched approximation without encountering exceptional con\-fig\-u\-rations. The presence of quenched chiral logarithms is studied under the tentative assumption of Goldstone Boson-like behaviour.}

\FullConference{31st International Symposium on Lattice Field Theory - LATTICE 2013\\
		July 29 - August 3, 2013\\
		Mainz, Germany}

\begin{document}

\section{Introduction}

For nearly 25 years of research on QCD, reconciliation of chiral symmetry and the lattice regularisation was considered a formidable problem. The Nielsen-Ninomiya no-go theorem prohibits local definitions of single chiral modes at finite cutoff, which reproduce the Dirac operator in the continuum limit.
Minimally doubled fermions comply with the no-go theorem by having two chiral modes, which are detached in the Brillouin zone, but still degenerate in the continuum limit. Their displacement defines an explicit violation of hypercubic symmetry, which entails counterterms \cite{1} with the same reduced symmetry.

Karsten-Wilczek fermions \cite{2} are a particular class of minimally doubled fermions with two residual zero modes from the original na\"\i ve fermion action. The Wilczek parameter $ \zeta $, which must satisfy $ |\zeta|>1/2 $, is by default fixed to $ 1 $. The full action \cite{1} reads

\small
\begin{eqnarray}
 S_{f,\alpha}^{KW}
 &=&
 \sum\limits_{x}\sum\limits_{\mu}\tfrac{1+d(g_0^2)\delta_{\mu\alpha}}{2a}
 \big(\overline{\psi}_{x} \gamma_{\mu} U_{\mu}(x) \psi_{x+\hat{\mu}}
 -\overline{\psi}_{x+\hat{\mu}} \gamma_{\mu} U_{\mu}^{\dagger}(x-\hat{\mu}) \psi_{x}\big)+
 \big(\overline{\psi}_{x} m_{0} \psi_{x}\big) \nonumber \\
 &-&
 \sum\limits_{\mu\neq\alpha}i\tfrac{\zeta}{2a}
 \big(\overline{\psi}_{x} \gamma_{\alpha} U_{\mu}(x) \psi_{x+\hat{\mu}}
 +\overline{\psi}_{x+\hat{\mu}} \gamma_{\alpha} U_{\mu}^{\dagger}(x-\hat{\mu}) \psi_{x}\big)+
 \big(\overline{\psi}_{x} \left(i\tfrac{3\zeta+c(g_0^2)}{a}\gamma_{\alpha}\right) \psi_{x}\big), \label{eq: Sf}\\
 S_{g,\alpha}^{KW}
 &=&
 \beta \sum\limits_{x}\sum\limits_{\mu<\nu}\big(1-\tfrac{1}{N_c}\mathrm{Re} \mathrm{Tr} P_{\mu\nu}(x)\big)(1+d_P(g_0^2)\delta_{\mu\alpha}), \label{eq: Sg}
\end{eqnarray}\normalsize
which includes three counterterms.
The zero modes are aligned at $ k_\alpha=0 $ and $ k_\alpha=\pi/a $ on the $ x_\alpha $-axis, which is commonly chosen as $ x_\alpha=x_0 $. The spinor field $ \psi(x) $ simultaneously contains two tastes with degenerate continuum limit, which are treated as the light quarks.
%
The Karsten-Wilzek term has non-singlet taste structure \cite{3}
and explicitly breaks $ x_\alpha $-reflection and charge symmetry, but is invariant under their product \cite{3,4}.

Point-split vector and axial currents are obtained with chiral Ward-Takahashi identities. Their conservation has been verified at 1-loop level \cite{1}. Counterterms which expli\-cit\-ly break hypercubic symmetry are indispensable for restoring isotropy to the continuum limit.
One relevant and two marginal operators share the Karsten-Wilczek term's symmetry. Renormalisation of the Karsten-Wilczek action at 1-loop level is covered in great detail in \cite{1}.
Mixing with these operators reflects in anisotropies in the fermionic self-energy,
\small
\begin{equation}
 \Sigma = \Sigma_{1} i\slashed{p} + \Sigma_{2}m_{0} + d_{1L}\,i(\gamma_{\alpha}\,p_{\alpha}) + c_{1L} \dfrac{i}{a}\,\gamma_{\alpha},
\end{equation}\normalsize
and in the fermionic contribution to the vacuum polarisation,
\small
\begin{equation}
 \left(p_{\mu}p_{\nu}(\delta_{\alpha\mu}+\delta_{\alpha\nu})-\delta_{\mu\nu}(p^2\delta_{\alpha\mu}\delta_{\alpha\nu}+p_{\alpha}^2)  \right)\times d_{P,\,1L}.
\end{equation}\normalsize
These anisotropies are removed by setting the coefficients to their $ 1 $-loop values,
\small
\begin{equation}
 c=c_{1L}=-29.5320\,C_{F}\,b,\ d=d_{1L}=-0.12554\,C_{F}\,b,\ d_{P}=-12.69766\,C_{2}\,b,\ b=\dfrac{g_{0}^2}{16\pi^2}.
\end{equation}\normalsize
The coefficients inherit the taste structure of the Karsten-Wilczek term:
\small
\begin{equation}
 c_{1L}(-\zeta)=-c_{1L}(\zeta),\ d_{1L}(-\zeta)=+d_{1L}(\zeta),\ d_{P,\,1L}(-\zeta)=+d_{P,\,1L}(\zeta).
\end{equation}\normalsize
If minimally doubled fermions are employed in numerical simulations, it is desirable to determine the coefficients non-perturbatively. Boosted perturbation theory \cite{5} employing Parisi's coup\-ling yields estimates (cf. table \ref{tab: boosted coefficients}), which are often close to non-perturbative values.
\begin{table}[hbt]
\center\footnotesize
 \begin{tabular}{|c|c|c|c|c|c|c|c|}
  \hline
  $ \beta $ & $ U_{0}^4 $ & $ c_{1L} $  & $ c_{BPT} $ & $ d_{1L} $ & $ d_{BPT} $ & $ d_{P,\,1L} $ & $ d_{P,\,BPT} $ \\
  \hline
  $ 6.0 $ & $ 0.594 $ & $ -0.249 $ & $ -0.420 $ & $ -0.00106 $ & $ -0.00179 $ & $ -0.0893 $ & $ -0.150 $ \\
  \hline
  $ 6.2 $ & $ 0.614 $ & $ -0.241 $ & $ -0.393 $ & $ -0.00103 $ & $ -0.00167 $ & $ -0.0865 $ & $ -0.141 $ \\
  \hline
 \end{tabular}
 \caption{Boosted 1-loop coefficients serve as starting point for non-perturbative determinations. Non-perturbative effects are estimated with the fourth root of the average plaquette, $ U_0=\sqrt[4]{\langle \sum_{\mu<\nu}P_{\mu\nu}\rangle} $. Numerical values for $ U_0^4 $ are taken from \cite{6}.}
 \label{tab: boosted coefficients}
 \vspace{-8pt}
\end{table}

\section{Non-perturbative renormalisation}

The violation of hypercubic symmetry in the Karsten-Wilczek action and its counter\-terms manifests itself as an anisotropy of the transfer matrix of QCD. 
Never\-the\-less, fully-tuned counterterm coefficients must minimise the degree of anisotropy which is observed at finite lattice spacing. 
Hence, the most straightforward strategy for non-perturbative tuning is a comparison of computations of correlation functions in different euclidean directions. Since the strength of anisotropies due to the action is a priori unclear, additional causes of anisotropy (e.g. $ L \neq T $) must be avoided.

\subsection{Numerical procedure}

\begin{table}[hbt]
\center\footnotesize
 \begin{tabular}{|c|c|c|c|c|c|c|c|}
  \hline
  $ \beta $ & $ a\,[fm] $ & $ r_0 $ & $ L $ & $ n_{cfg} $ & $ m_0\,(\times100) $ & $ c $ & $ d\,(\times1000) $ \\
  \hline
  $ 6.0 $  & $ 0.093 $ & $ 5.368 $ & $ 32 $ & $ 100 $ & $ 2,3,4,5 $ &
  $ [-1.2,+0.3]  $ & $ 0.0 $ \\
  $ 6.0 $  & $ 0.093 $ & $ 5.368 $ & $ 32 $ & $ 100 $ & $ 1,2,3,4,5 $ &
  $ [-0.65,-0.20] $ & $ [-8,+2] $ \\
  $ 6.0 $  & $ 0.093 $ & $ 5.368 $ & $ 48 $ & $  40 $ & $ 2 $ &
  $ [-0.65,+0.0]  $ & $ 0.0 $ \\
  \hline
  $ 6.2 $  & $ 0.068 $ & $ 7.360 $ & $ 32 $ & $ 100 $ & $ 1,2,3,4,5 $ & 
  $ [-0.65,-0.20]  $ & $ [-8,+4] $ \\
  $ 6.2 $  & $ 0.068 $ & $ 7.360 $ & $ 48 $ & $  40 $ & $ 2 $ &
  $ [-0.65,+0.0]  $ & $ 0.0 $ \\
  \hline
  $ 5.8 $  & $ 0.136 $ & $ 3.668 $ & $ 32 $ & $ 100 $ & $ 2 $ & 
  $ [-0.65,+0.0]  $ & $ [0,+2] $ \\
  \hline
 \end{tabular}
 \caption{Symmetric lattices ($ T=L $) were used for studies of the anisotropy. The parameter $ c $ is varied with smaller step size close to the estimates from boosted perturbation theory. The scale was fixed using the Sommer parameter according to \cite{7}.}
 \label{tab: lattice parameters}
 \vspace{-4pt}
\end{table}\normalsize
In the quenched approximation, $ d_P $ equals zero due to the absence of virtual quark loops\footnote{
In full QCD, $ d_P $ is fixed by restoring the isotropy of the plaquette at fixed $ c,d $ (cf. \cite{1}). 
} 
and four-dimensional parameter space is spanned by $ \{\beta,m_0,c,d\} $. Simulations are performed on symmetric lattices ($ L=T $) using the temporal Karsten-Wilczek action ($ x_\alpha=x_0 $) with default Wilczek parameter ($ \zeta=+1 $). 
Gaussian smearing \cite{8} at the source is combined with local and smeared sink operators using HYP-smeared \cite{9} link variables. Here, we restrict the discussion to pseudoscalar correlation functions.

The relevant parameter $ c $ is varied at fixed coupling $ \beta $ and quark mass $ m_0 $ in order to establish an smooth relation between hadronic quantities and renormalisation coefficients. The small size of  $ d $ in perturbation theory suggests that its influence is mild; hence, we set $ d=0 $ initially. The difference of pseudoscalar fit masses of both directions, the mass anisotropy,
\small
\begin{equation}
 \Delta(M_{PS}^2)=(M_{PS}^{x_0})^2-(M_{PS}^{x_3})^2,
 \label{eq: mass anisotropy}
\end{equation}\normalsize
is used as a tuning criterion for $ c $ at a fixed value of $ d $ and several values of the the bare quark mass. Finally, effects due to the variation of $ d $ are studied.


\subsection{Determination of the pseudoscalar mass}


\begin{figure}[htb]
 \begin{picture}(360,105)
  \put( 15  , 0.0){\includegraphics[height=40mm]{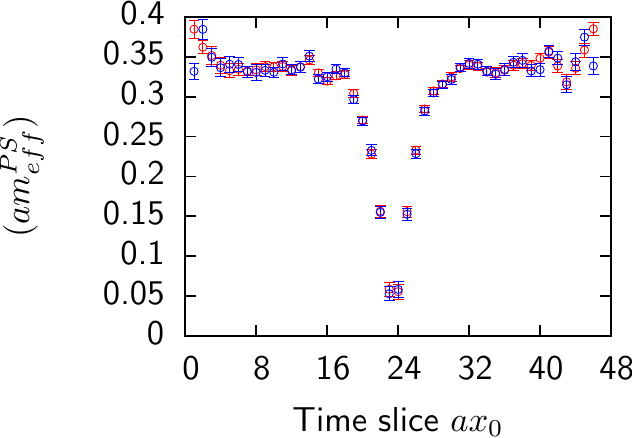}}
  \put(215.0, 0.0){\includegraphics[height=40mm]{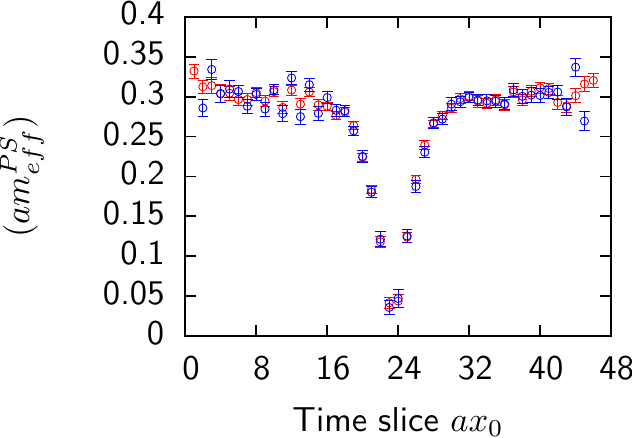}}
 \end{picture}
 \vspace{-4pt}
 \caption{ Effective mass plots ($ \beta=6.0,\ L=48,\ m_0=0.02,\ d=0 $) using the ``log'' mass exhibit isolated plateaus of forward (e.g. 8-16) and backward (e.g. 32-40) states. Local (red) and smeared sink (blue) are in good agreement. The left plot shows $ c=0.0 $ and the right plot shows $ c=-0.45 $.
}
 \label{fig: x0 effective mass}
 \vspace{-4pt}
\end{figure}
The Karsten-Wilczek term explicitly breaks $ T $-symmetry. Thus, it is conceivable that forward and backward propagating states in the $ x_0 $-direction are not degenerate. Hence, $ x_0 $-correlation functions must not be symmetrised. Forward and backward states are separated when the effective mass is obtained as a logarithm of the correlation function,
\small
\begin{equation}
  m_{log}(t)=\log\frac{\mathcal{C}(t)}{\mathcal{C}(t+1)}.
  \label{eq: log mass}
  \vspace{-4pt}
\end{equation}\normalsize
The single exponential does not provide a good description of the data around $ T/2 $ (cf. figure \ref{fig: x0 effective mass}). Fits to the correlation function probe forward and backward states independently:
\small
\begin{equation}
 C_{PS}(t)\equiv A_{f}e^{-m_{f}t}+A_{b} e^{-m_{b}(T-t)}.
 \vspace{-4pt}
\end{equation}\normalsize
The 4-parameter fit extracts forward and backward masses as independent parameters. ``Log'' mass plateaus agree within errors with definitions of the ``cosh'' mass.
No numerical evidence of broken \textit{T}-symmetry was found regardless of $ c $ within $ 1\,\sigma $ (cf. table \ref{tab: forward and backward masses}).
\begin{table}[hbt]
\center\footnotesize
 \begin{tabular}{|c|c|c|c|c|c|c|}
  \hline
  $ c  $   & $ m_f $ (SL) & $m_b $ (SL) & $m_s $ (SL) & $ m_f $ (SS) & $m_b $ (SS) & $m_s $ (SS) \\
  \hline
  $ +0.0 $  & $ 0.3373(19) $ & $ 0.3370(18) $ & $ 0.3372(14) $ & $ 0.3355(17) $ & $ 0.3375(18) $ & $ 0.3364(13) $ \\
  $ -0.45 $ & $ 0.3036(25) $ & $ 0.3027(18) $ & $ 0.3029(15) $ & $ 0.3014(18) $ & $ 0.3008(15) $ & $ 0.3010(13) $ \\
  \hline
 \end{tabular}
 \caption{Forward and backward fit masses in the $ x_0 $-direction ($ \beta=6.0,\ L=48,\ m_0=0.02,\ d=0 $) with local and smeared sink agree within $ 1 $-$ 2\,\sigma $. A tentative ``cosh'' mass fit is consistent within $ 1\,\sigma $.
}
 \label{tab: forward and backward masses}
 \vspace{-8pt}
\end{table}


\begin{table}[hbt]
\center\footnotesize
 \begin{tabular}{|c|c|c|c|c|}
  \hline
  $ c  $    & $ M $ (SL, $ [12,23] $) & $ M $ (SL, $ [20,23] $) & $ M $ (SS, $ [12,23] $) & $ M $ (SS, $ [20,23] $) \\
  \hline
  $ +0.0 $  & $ 0.2823(9) $ & $ 0.2831(17) $ & $ 0.2963(10) $ & $ 0.2917(20) $ \\
  $ -0.45 $ & $ 0.2961(11) $ & $ 0.2973(21) $ & $ 0.3046(10) $ & $ 0.3048(23) $ \\
  \hline
 \end{tabular}
 \caption{Fit masses of smeared-local and smeared-smeared correlation functions in the $ x_3 $-direction ($ \beta=6.0,\ L=48,\ m_0=0.02,\ d=0 $) agree within $ 3\,\sigma $.
}
 \label{tab: x_3 masses}
 \vspace{-4pt}
\end{table}

\begin{figure}[htb]
 \begin{picture}(360,105)
  \put( 15  , 0.0){\includegraphics[height=40mm]{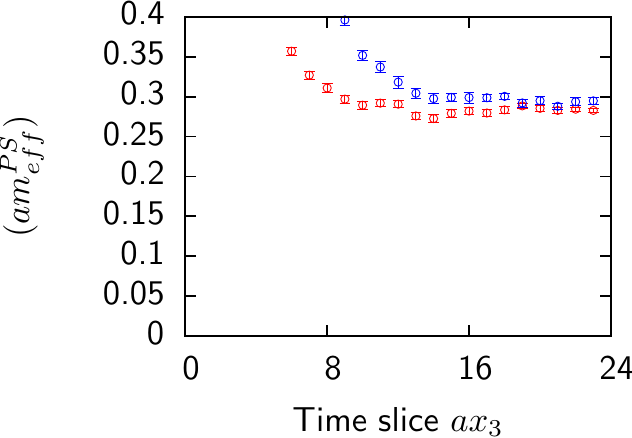}}
  \put(215.0, 0.0){\includegraphics[height=40mm]{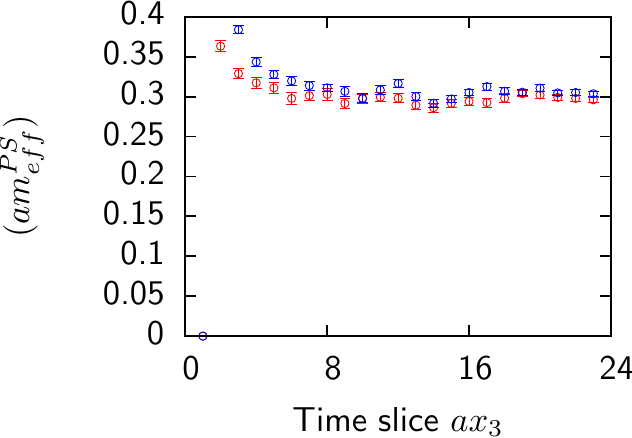}}
 \end{picture}
 \vspace{-4pt}
 \caption{Effective mass plots ($ \beta=6.0,\ L=48,\ m_0=0.02,\ d=0 $) in the $ x_3 $-direction using the ``cosh'' mass are calculated from symmetrised correlation functions with local (red) and smeared (blue) sink. Plateaus are considerably shorter for $ c=0 $ (left plot) than for $ c=-0.45 $ (right plot).
}
 \label{fig: x3 effective mass}
 \vspace{-8pt}
\end{figure}
Effective masses in the $ x_3 $-direction (cf. figure \ref{fig: x3 effective mass}) are computed from correlation functions, which were symmetrised over forward and backward propagating states. Excited state contributions persist longer than in figure \ref{fig: x0 effective mass}. The plateaus are more extended in the vicinity of $ c_{BPT} $ (cf. table \ref{tab: boosted coefficients}). Figure \ref{fig: x3 effective mass} demonstrates that effective masses of $ x_3 $-correlation functions with local and smeared sink interpolators reach $ 1 $-$ 2\,\sigma $ level agreement only after $ 16 $-$ 18 $ time slices at $ c=0.0 $ (cf. table \ref{tab: x_3 masses}). Therefore, this analysis of the mass anisotropy with $ L=32 $ (cf. section \ref{sec: minimisation of the anisotropy}) uses only local sinks.

\section{Numerical results}
 \vspace{-8pt}
\subsection{Minimisation of the anisotropy}
\label{sec: minimisation of the anisotropy}

\begin{figure}[hbt]
 \begin{picture}(360,105)
  \put(15.0, 0.0){\includegraphics[height=40mm]{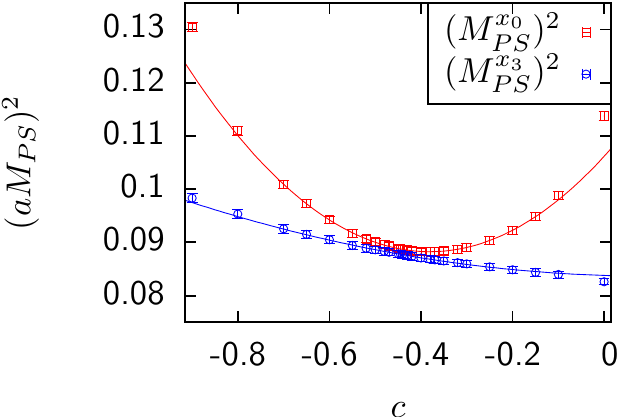}}
  \put(215.0, 0.0){\includegraphics[height=40mm]{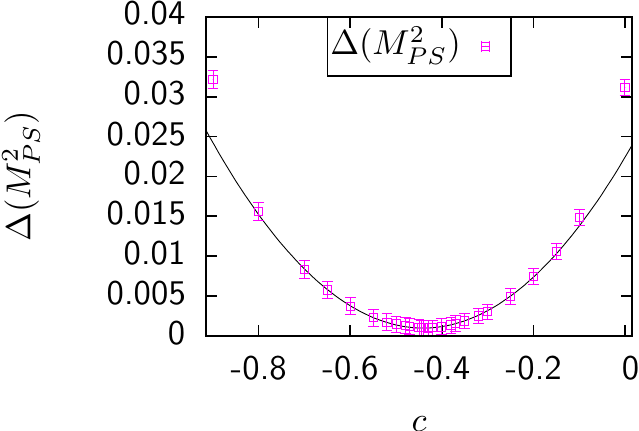}}
 \end{picture}
 \vspace{-8pt}
 \caption{Fit masses ($ \beta=6.0,\ L=32,\ m_0=0.02,\ d=0.0 $) are interpolated in $ c \in [-0.65,-0.25] $. The minimum of $ \Delta(M_{PS}^2) $ as a function of $ c $ (right plot) is shallow with respect to statistical errors.
}
 \label{fig: interpolation}
 \vspace{-4pt}
\end{figure}
\begin{figure}[htb]
\begin{picture}(360,105)
  \put(15.0, 0.0){\includegraphics[height=40mm]{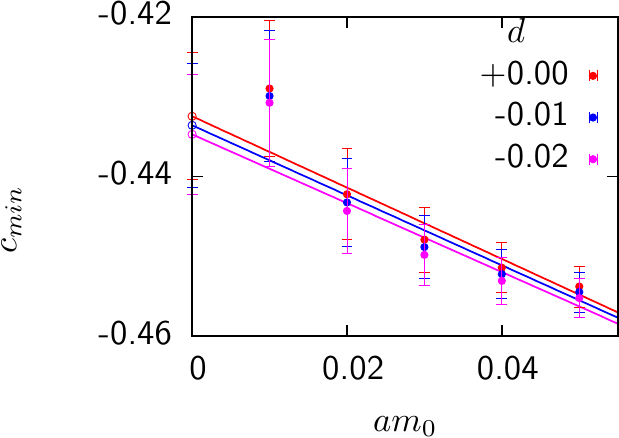}}
  \put(215.0, 0.0){\includegraphics[height=40mm]{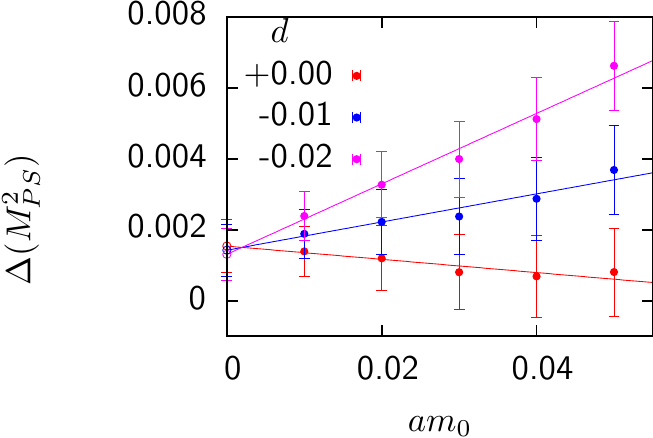}}
 \end{picture}
 \vspace{-8pt}
 \caption{Dependence on $ d $ of $ c_{min} $ ($ \beta=6.0,\ L=32 $) cannot be resolved (left plot). The minimal fit mass anisotropy is consistent with zero on a $ 2\,\sigma $ level (right plot).
}
 \label{fig: extrapolation}
  \vspace{-4pt}
\end{figure}

The minimisation of eq. (\ref{eq: mass anisotropy}) as a function of $ c $ and $ d $ defines the renormalisation condition. The squared fit masses $ (M_{PS}^{x_\mu})^2 $ with local sinks are interpolated as functions of $ c $ (cf. figure \ref{fig: interpolation}). The interpolations are directly subtracted and the minimum is computed,
\small
\begin{equation}
 c_{min}=-\tfrac{(a_1^{x_0}-a_1^{x_3})}{2(a_2^{x_0}-a_2^{x_3})},\quad (M_{PS}^{x_\mu})^2=a_0^{x_\mu}+a_1^{x_\mu}\,c+a_2^{x_\mu}\,c^2.
 \vspace{-4pt}
\end{equation}\normalsize
$ c_{min} $ is extrapolated (cf. figure \ref{fig: extrapolation}) in the quark mass $ m_0 $ with a linear and a quadratic ansatz, which agree at $ 1 $-$ 2\,\sigma $ level. The error is dominated by the lightest quark mass.

\begin{table}[hbt]
\center\footnotesize
 \begin{tabular}{|c|c|c|c|c|c|}
  \hline
  $ \beta $ & $ c_{BPT}  $ & $ c_{min} $ (lin.) & $ \Delta(M_{PS}^2) $ (lin.) & $ c_{min} $ (quad.) & $ \Delta(M_{PS}^2) $ (quad.) \\
  \hline
  $ 6.0 $ &  $ -0.420 $ & $ -0.432(08)^{\text{stat}} $ & $ 0.0015(07)^{\text{stat}}  $ & $ -0.418(13)^{\text{stat}} $ & $ 0.0019(08)^{\text{stat}}  $ \\
  \hline
  $ 6.2 $ &  $ -0.393 $ & $ -0.413(16)^{\text{stat}} $ & $ -0.0015(12)^{\text{stat}} $ & $ -0.414(35)^{\text{stat}} $ & $ -0.0006(13)^{\text{stat}} $ \\
  \hline
 \end{tabular}
 \caption{The different $ c_{min} $ ($ L=32,\ d=0.0 $) from linear and quadratic extrapolations in the quark mass ($ am_0\in[0.1,0.5] $) agree on $ 1 $-$ 2\,\sigma $ level. The mass anisotropy scatters around $ 0 $ within  $ 1 $-$ 2\,\sigma $.
}
 \label{tab: cmin}
 \vspace{-4pt}
\end{table}
We conclude that use of the parameter estimates ( $ c_{BPT} $, $ d_{BPT} $) from boosted perturbation theory removes the mass anisotropy within our statistic and systematic accuracy. However, careful study of additional observables \cite{10} indicates slightly different values ($ c(\beta=6.0)=-0.45(1) $, $ c(\beta=6.2)=-0.40(1) $), which we use in studies of the tuned action.

\subsection{Simulations with the tuned Karsten-Wilczek action}

\begin{figure}[hbt]
 \begin{picture}(360,105)
  \put(020.0, 0.0){\includegraphics[height=40mm]{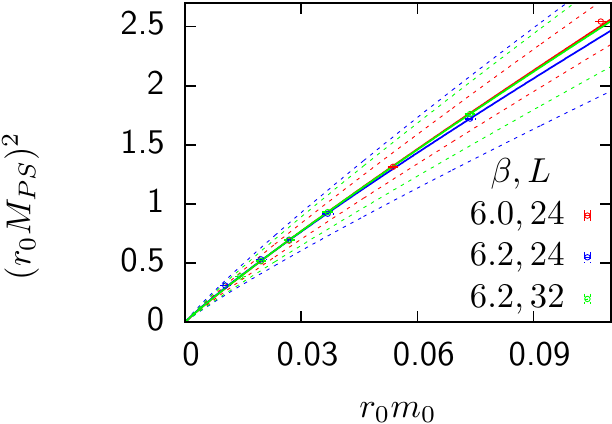}}
  \put(245.0, 0.0){\includegraphics[height=40mm]{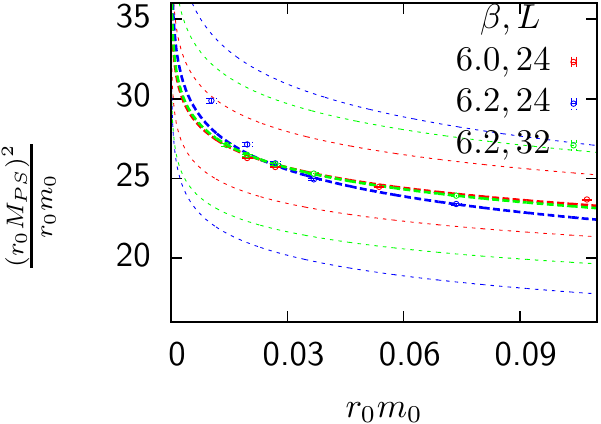}}
 \end{picture}
 \vspace{-4pt}
 \caption{The pseudoscalar mass at $ \beta=6.0 $ and $ \beta=6.2 $ agrees well (left plot). The ratio {\small $ {(r_0 M_{PS})^2}/{(r_0 m_0)} $} shows finite volume effects and quenched chiral logarithms.
}
 \label{fig: chiral limit}
 \vspace{-4pt}
\end{figure}

\begin{table}[hbt]
\center\footnotesize
 \begin{tabular}{|c|c|c|c|c|c|c|c|c|}
  \hline
  $ \beta $ & $ c  $ & $ d $ & $ m_0\,(\times1000) $ & $ (r_0 m_0) $ & $ (r_0 M_{PS})^2 $ & $ (M_{PS})^2\ [MeV] $ & $ \tfrac{(r_0 M_{PS})^2}{(r_0 m_0)} $ \\
  \hline
  $ 6.0 $ &  $ -0.45 $ & $ -0.001 $ & $ 20 $   & $ 0.107 $ & $ 1.595(4)
 $ & $ 629(2) $ & $ 23.7(1) $  \\
  $ 6.0 $ &  $ -0.45 $ & $ -0.001 $ & $ 10 $   & $ 0.054 $ & $ 1.147(4)
 $ & $ 452(2) $ & $ 24.5(2) $  \\
  $ 6.0 $ &  $ -0.45 $ & $ -0.001 $ & $ 5 $    & $ 0.027 $  & $ 0.831(5) $ & $ 328(2) $ & $ 25.7(3) $  \\
  $ 6.0 $ &  $ -0.45 $ & $ -0.001 $ & $ 3.65 $ & $ 0.020 $  & $ 0.718(5) $ & $ 283(2) $ & $ 26.3(4) $  \\
  \hline
  $ 6.2 $ &  $ -0.40 $ & $ -0.001 $ & $ 20^* $   & $ 0.147^* $ & $ 1.834(9)^* $ & $ 724(3)^* $ & $ 22.8(2)^* $  \\
  $ 6.2 $ &  $ -0.40 $ & $ -0.001 $ & $ 10 $   & $ 0.074 $ & $ 1.327(8) $ & $ 524(3) $ & $ 23.9(3) $  \\
  $ 6.2 $ &  $ -0.40 $ & $ -0.001 $ & $  5 $   & $ 0.037 $ & $ 0.965(8) $ & $ 381(3) $ & $ 25.3(4) $  \\
  $ 6.2 $ &  $ -0.40 $ & $ -0.001 $ & $ 3.65 $ & $ 0.027 $ & $ 0.834(9) $ & $ 329(3) $ & $ 25.9(5) $  \\
  $ 6.2 $ &  $ -0.40 $ & $ -0.001 $ & $ 2.66 $ & $ 0.020 $ & $ 0.720(9) $ & $ 284(4) $ & $ 26.5(7) $  \\
  $ 6.2 $ &  $ -0.40 $ & $ -0.001 $ & $ 1.94 $ & $ 0.014 $ & $ 0.622(10) $ & $ 245(4) $ & $ 27.1(9) $  \\
  $ 6.2 $ &  $ -0.40 $ & $ -0.001 $ & $ 1.41^* $ & $ 0.010^* $ & $ 0.557(18)^* $ & $ 219(7)^* $ & $ 29.9(19)^* $  \\
  \hline
 \end{tabular}
 \caption{The tuned action is simulated on lattices with $ T=48 $. The spatial extent is $ L=24 $ for $ \beta=6.0 $ and $ L=32 $ for $ \beta=6.2 $. Two parameter sets (marked with ``$ ^* $'') have only $ L=24 $.
}
 \label{tab: simulation parameters}
 \vspace{-4pt}
\end{table}
The action with tuned parameters is applied to a study of the spectrum of light pseudoscalar mesons (cf. table \ref{tab: simulation parameters}). Ground state masses below $ 250\,MeV $ are achieved without encountering exceptional configurations. Since the squared ground state mass is approximately linear in the quark mass (cf. figure \ref{fig: chiral limit}), it is tentatively extrapolated like a Goldstone boson including quenched chiral logarithms \cite{10},
\small
\begin{equation}
 (r_{0}\,M_{PS})^2 = (r_{0}\,B_{0})(r_{0}\, m_{0})\left((1-\delta) - \delta \log (m_{0}/r_{0})\right).
\end{equation}\normalsize

However, the separation of chiral logarithms from effects due to finite volume or higher chiral orders is difficult. With the enlarged volume, the statistical error of $ \delta $ decreases and agreement between different lattice spacings is improved considerably. 

We obtain the estimate $ 0.10(1) \leq \delta \leq 0.16(5)) $ and find consistency of $ \delta $ between different lattice spacings and different volumes within $ 2\,\sigma $.

\section{Conclusions}

The first simulations with minimally doubled fermions in the quenched approximation have been performed with various volumes and different lattice spacings (cf. table \ref{tab: lattice parameters}). Pseudoscalar correlation functions do not show any numerical evidence of \textit{T}-parity violation. This surprising result is currently under scrutiny \cite{10}. Anisotropies of the pseudoscalar  masses are applied to determine $ c $ non-perturbatively. Results are largely insensitive to $ d $ and agree well with estimates from boosted perturbation theory (cf. table \ref{tab: cmin}). However, separate methods for obtaining $ c $ and $ d $ with reduced errors are still desirable \cite{10}.

The tuned action (cf. table \ref{tab: simulation parameters}) is used in studies of light pseudoscalar mesons ($ M_{PS}\lesssim250\,MeV $) without exceptional configurations. After taking quenched chiral logarithms into account, 
 the ground state is consistent with a Goldstone boson. This remarkable result requires a detailed study of the nature of the pseudoscalar ground state \cite{10}.

\vskip1em
\textbf{Acknowledgements}: {The speaker thanks Sinya Aoki and Michael Creutz for invaluable discussions. This work was supported by Deutsche Forschungsgemeinschaft (SFB 1044), Gesellschaft f\"ur Schwerionenforschung GSI, the Research Center ``Elementary Forces \& Mathematical Foundations'' (EMG), Helmholtz Institute Mainz (HIM), Deutscher Akademischer Austauschdienst (DAAD) and the Japanese Ministry of Education, Culture, Sports and Technology (MEXT). Simulations have been performed on the cluster ``Lilly'' at the Institue for Nuclear Physics, Univ. of Mainz. We thank C. Seiwerth for technical support.}

\end{document}